\documentclass[reprint,superscriptaddress,amsmath,amssymb,aps,showpacs,prl,twocolumn]{revtex4-1}

\usepackage[T1]{fontenc}
\usepackage[utf8]{inputenc}
\usepackage{graphicx}
\usepackage{bm}
\usepackage{epsfig}
\usepackage{amssymb}
\usepackage{amsfonts}
\usepackage{braket}
\usepackage{color}
\usepackage{epstopdf}
\usepackage{hyperref}
\usepackage{float}
\usepackage{bibentry}
\usepackage[caption=false]{subfig}

\newcommand{\ba}{\begin{eqnarray}}
\newcommand{\ea}{\end{eqnarray}}
\newcommand{\bd}{\begin{displaymath}}
\newcommand{\ed}{\end{displaymath}}
\renewcommand{\v}[1]{{\bf #1}}
\newcommand{\bpm}{\begin{pmatrix}}
\newcommand{\epm}{\end{pmatrix}}

\graphicspath{{figure/}}

\begin{document}
\title{Formation of a Topological Monopole Lattice and its Dynamics in \\ Three-dimensional Chiral Magnets}
\author{Seong-Gyu Yang}
\affiliation{Department of Physics, Sungkyunkwan University, Suwon 16419, Korea}
\author{Ye-Hua Liu}
\affiliation{Theoretische Physik, ETH Zurich, 8093 Zurich, Switzerland}
\author{Jung Hoon Han}\email{hanjh@skku.edu}
\affiliation{Department of Physics, Sungkyunkwan University, Suwon 16419, Korea}

\date{\today}

\begin{abstract}
Topologically protected swirl of the magnetic texture known as the Skyrmion has become ubiquitous in both metallic and insulating chiral magnets. Meanwhile the existence of its three-dimensional analogue, known as the magnetic monopole, has been suggested by various indirect experimental signatures in MnGe compound. Theoretically, Ginzburg-Landau arguments in favor of the formation of a three-dimensional crystal of monopoles and anti-monopoles have been put forward, however no microscopic model Hamiltonian was shown to support such a phase. Here we present strong numerical evidence from Monte Carlo simulations for the formation of a rock-salt crystal structure of monopoles and anti-monopoles in short-period chiral magnets. Real-time simulation of the spin dynamics suggests there is only one collective mode in the monopole crystal state in the frequency range of several GHz for the material parameters of MnGe.
\end{abstract}


\maketitle
Attempts to view building blocks of nature as topologically protected objects such as knots have been a fascinating feature of modern physical science. The knot theory of atoms advanced by Thomson (Lord Kelvin) \cite{thompson} was re-incarnated by Skyrme as a topological quantum theory of hadrons in the early 60s \cite{skyrme}. Although the topological interpretation of quantum numbers for elementary particles may not have been universally accepted in sub-atomic physics, the beauty of the idea has remained well within the physics community and, quite recently, found immense physical realization in several kinds of magnetic materials ranging from B20 metallic magnetic compounds \cite{pfleiderer09,YiSuDo,tokura10}, atomically-thin magnetic layers \cite{wiesendanger}, multiferroic insulators \cite{tokura-MF-skyrmion}, to ferroelectrics \cite{ferroelectric_Sk}. The form of the topological matter, now called the Skyrmion lattice, bears excellent resemblance to the well-known vortex lattice in type-II superconductors \cite{han} and shares the character of two dimensionality, extending its existence in a columnar fashion when the host material in which it is formed is three-dimensional. The topological vortex and Skyrmion lattices have both been observed successfully through the visualization technique of Lorentz microscopy \cite{tonomura,tokura10}.

\begin{figure*}[t]
\includegraphics[width=1.0\textwidth]{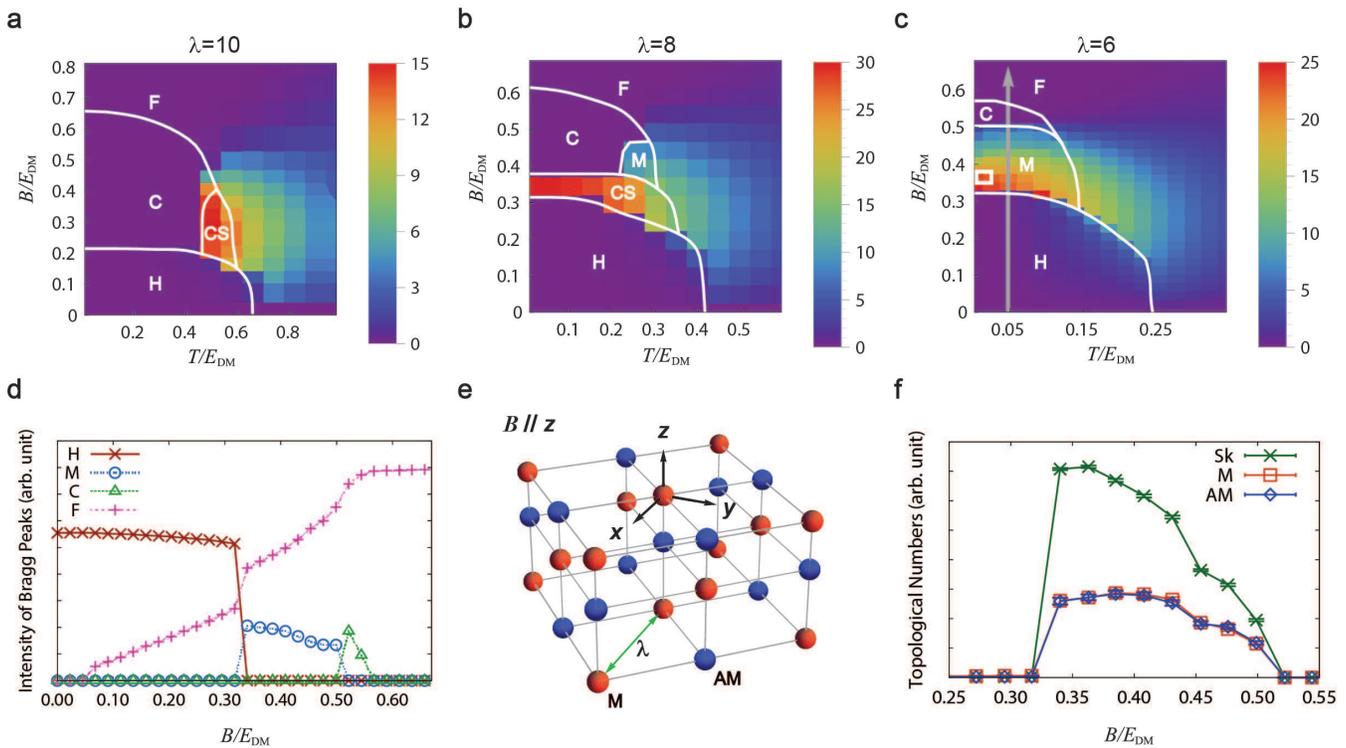}
\caption{Phase diagrams for increasing strengths of the Dzyaloshinskii-Moriya interaction. The wavelength of spin modulations is set to (a) $\lambda=10$, (b) $\lambda=8$, and (c) $\lambda=6$, respectively, with the lattice constant equal to unity. Color plots indicate the integrated Skyrmion density $ (1/4\pi) \int dx dy \, \v n \cdot (\partial_x \v n \times \partial_y \v n)$ within the $xy$-plane, further averaged over all the layers in the $z$-direction. An energy unit $E_{\rm DM}=D^2/J$ is used to normalize the temperature and magnetic-field scales. Acronyms indicate helical (H), columnar Skyrmion (CS), conical (C), ferromagnetic (F), and monopole-anti-monopole crystal (M) phases, respectively. (d) Intensity of Bragg peaks corresponding to different phases along the gray arrow in (c). Sharp rise or fall in the intensity occurs at all phase boundaries. (e) Rock-salt structure of the monopole-anti-monopole crystal, at the location of the empty white square in the phase diagram (c). Red and blue balls represent monopoles and anti-monopoles, respectively.  (f) Variation of the Skyrmion (Sk), monopole (M), and anti-monopole (AM) topological numbers with increasing magnetic field at $\lambda=6$ along the gray arrow in (c). Monopoles and anti-monopoles always occur in equal numbers. Magnetic field dependences of both topological numbers are similar. Statistical variations are less than the size of the symbols.}
\label{fig:Fig1}
\end{figure*}

In higher dimensions one is granted the exciting opportunity to create, observe, and manipulate topological objects not permitted in lower dimensions. For example in three-dimensional Heisenberg magnets with a local unit magnetization vector $\v n_{\v r}$, localized objects known as monopoles and anti-monopoles with integer topological numbers may be formed. The number characterizing the object's topology is
\begin{align}
\frac{1}{4\pi} \int_{S} du dv ~ \v n_{\v r} \cdot \left( \frac{\partial \v n_{\v r} }{\partial u} \times \frac{\partial \v n_{\v r} }{\partial v} \right)  =m \in {\rm integer} ,
\end{align}
for an integration domain $\int_{S}$, parameterized by $(u,v)$, enclosing the monopole center $\v R$. Due to the high energetic cost of creating monopoles in isolation, they must be created in reality as a monopole-anti-monopole (MAM) pair, or as a crystal of such pairs forming a new kind of topological lattice unique to three dimensions. Such possibility was suggested theoretically \cite{Binz-Vishwanath} some time ago following the intriguing discovery of diffuse neutron scattering peaks in MnSi under high pressure \cite{pfleiderer-Bragg-peak}, and was further corroborated in the calculation of Ref.~\onlinecite{Park-Han}. Several recent Hall and Nernst effect experiments on MnGe crystal have boosted our confidence that such three-dimensional topological phase has already been realized in nature even at ambient pressure \cite{tokura-MnGe-Hall,tokura-MnGe-Nernst}, while a more direct visual confirmation, analogous to the Lorentz microscopy imaging of the the two-dimensional Skyrmion lattice, may still be forthcoming \cite{3D-imaging}. A pressing issue for theory at this stage would be whether a simple magnetic model Hamiltonian supporting the MAM crystal phase can be written down, and used to address various aspects of the monopole dynamics.

We show in the present article that a well-known three-dimensional model of Heisenberg spins interacting through short-range ferromagnetic and Dzyaloshinskii-Moriya (DM) exchanges of strengths $J$ and $D$, respectively, supports the MAM phase at a sufficiently short wavelength $\lambda/a \sim  J/D$, where $a$ is the lattice constant. The same model is known to give rise to the columnar Skyrmion (CS) lattice phase within the so-called A-phase region of the phase diagram \cite{Lars}. The small pocket occupied by the A-phase expands to cover the entire low-temperature regime as the thickness of the sample is reduced below the helical wavelength \cite{tokura_thinfilm}. We show, through extensive Monte Carlo (MC) simulations of the same three-dimensional lattice model, that at shorter periods of the helix the A-phase gives way to the MAM crystal phase. The MAM phase (which we will also call the M-phase)  initially occurs around the high-temperature region just below the magnetic order and expands to cover the low-temperature region at a shorter helix period. In order to make the extensive three-dimensional simulation feasible we have developed an efficient Monte Carlo algorithm based on graphics processing units (GPUs) as explained in the supplemental material (SM). Subsequent to the identification of the M-phase we carry out the Landau-Lifshitz-Gilbert (LLG) simulation of the real-time spin dynamics in an effort to uncover the low-energy collective modes. Only one such mode could be identified well below the exchange energy scale $J$ for the MAM crystal.

\begin{figure*}[t]
\includegraphics[width=1.0\textwidth]{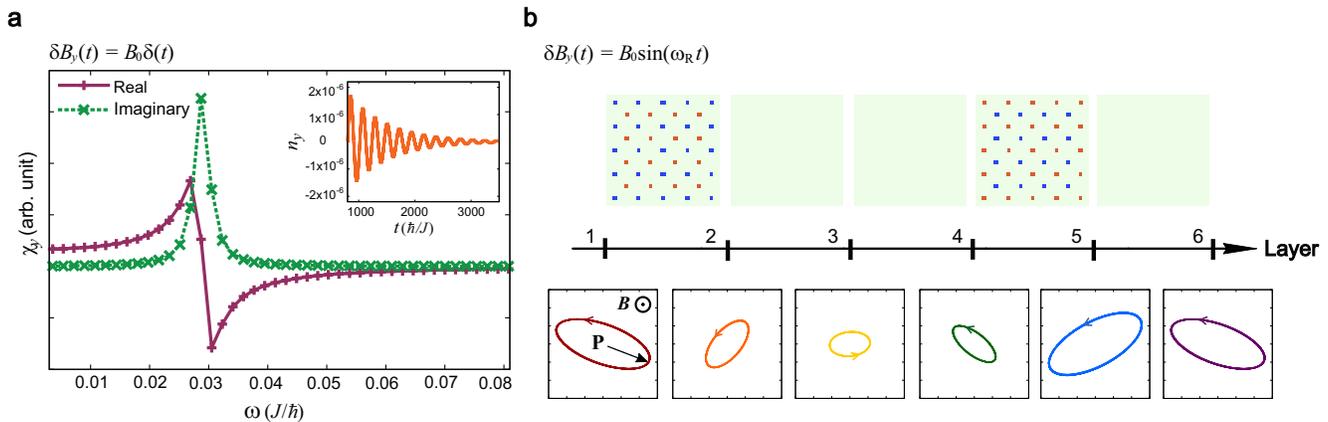}
\caption{Collective mode of the topological lattice. (a) Real and imaginary parts of the Fourier spectrum $\chi_y (\omega) = \int_0^\infty e^{i\omega t}n_y (t) dt$, which characterizes the response of the averaged magnetization along the $y$-direction $n_y (t)$, to the pulsed field $B_0 \delta(t)$ applied along the $y$-axis, $B_0/E_{\rm DM} = 0.0023$. Resonance is observed around the frequency $\omega_\mathrm{R} \sim0.03J/\hbar$. Inset shows the time-domain dynamics of $n_y (t)$. (b) Oscillation trajectories of the $\v P$-vector (see text for definition) in several layers along the $z$-direction, under the periodic pulse $\delta B_y (t) = B_0 \sin (\omega_{\rm R} t)$, $B_0/E_{\rm DM} = 0.023$. The sense of rotation of the $\v P$-vector is shown with arrows. Monopoles (red squares) and anti-monopoles (blue squares) are located at the dual cubic lattice positions and appear between layers 1 and 2, and also between layers 4 and 5. The whole structure, both the monopole locations and the $\v P$-vector trajectories, is repeated with period 5 along the $z$-direction. In-plane monopole spacing is 6. Simulations of the Landau-Lifshitz-Gilbert equation were performed on $L=30$ cubic lattice.}
\label{fig:Fig2}
\end{figure*}

Buhrandt and Fritz \cite{Lars} were the first to explicitly demonstrate the existence of the A-phase, nowadays identified with the CS phase, in their MC calculation for a three-dimensional model of chiral magnets. A phase diagram similar to theirs is shown in Fig.~\ref{fig:Fig1}(a) (we use $\lambda=10$ instead of $\lambda\sim$15 as in their work). From there, we reduce the helix period $\lambda$ by tuning the $D/J$ ratio and observe the evolution of the phase diagram. Throughout the simulation the Zeeman field is applied along the positive $z$-direction. As shown in Fig.~\ref{fig:Fig1}(b) for $\lambda=8$, the CS phase becomes stable even at very low temperatures where there used to be either the spiral or conical phase. One also starts to see the appearance of the M-phase in a small pocket that used to be occupied by the A-phase. For an even shorter wavelength $\lambda=6$, the low-temperature CS phase entirely gives way to the M-phase. As seen in Fig.~\ref{fig:Fig1}(c), a succession of phase transitions, from helical (H) to M then followed by conical (C) and ferromagnetic (F) phases, occur with increasing magnetic field at a fixed, low temperature. In producing various phase diagrams we adopted reduced units, in which both the temperature $T/E_{\rm DM}$ and the magnetic field $B/E_{\rm DM}$ are made dimensionless in terms of the energy scale $E_{\rm DM} = D^2/J$. Phase boundaries were efficiently identified by tracking the intensities of the Bragg peaks for each expected spin pattern. For the MAM crystal state we look for four pairs of Bragg peaks expected in the face-centered cubic (fcc) structure \cite{Binz-Vishwanath,Park-Han}, and for the helical phase the corresponding Bragg peaks occur along the (111) direction. Figure~\ref{fig:Fig1}(d) shows the evolution of H, M, and C phases in terms of their respective Bragg intensities. A clear phase boundary can be read off from the rise and fall of each intensity as a function of the magnetic field strength. MC calculations of thermodynamic quantities such as the specific heat and magnetic susceptibility offered compatible phase boundaries.

In the theoretical literature several possibilities for the monopole crystal structure were proposed, including the simple cubic, face-centered cubic, and body-centered cubic structures \cite{Binz-Vishwanath,Park-Han}. Among these proposals the one realized by the microscopic model we have adopted is the fcc structure. We conclude this by identifying monopole and anti-monopole locations from the MC-generated spin configuration (see SM for details of the monopole calculation) and plotting them in real space. As shown in Fig.~1(e), monopoles and anti-monopoles respectively form a fcc lattice and they together constitute a rock-salt structure. The monopole-monopole separation is equal to $\lambda$ in the plane orthogonal to the Zeeman field and slightly less along the field direction. Bragg-peak analysis of the spin structure in the M-phase also yields four non-coplanar peaks at $(2\pi/\lambda)(2,0,12/5)$, $(2\pi/\lambda)(-2,0,12/5)$, $(2\pi/\lambda)(0,2,-12/5)$ and $(2\pi/\lambda)(0,-2,-12/5)$, in conformity with the fcc structure. We have established that both the Skyrmion number and the monopole number rise and disappear at the same critical fields, i.e. they exist over the same field ranges $B_\mathrm{c_1} \le B \le B_\mathrm{c_2}$ as shown in Fig.~\ref{fig:Fig1}(f).

Low-frequency collective modes of the two-dimensional Skyrmion lattice are well understood theoretically \cite{mochizuki,yehua_han,Batista} and demonstrated by microwave experiments \cite{Wang,Tokura_resonance,Tokura_observation}. They arise from either the breathing mode (when Skyrmions expand and contract periodically) or the small cyclotron-like motion of the center of mass of the Skyrmions. With the characteristic exchange energy $J\sim 1$~meV and the frequency $f=J/h \sim 240$~GHz in most chiral magnets \cite{mochizuki,Wang}, collective modes have been found at a small fraction of this frequency around a few GHz \cite{Tokura_resonance,Tokura_observation}. Intuitively one expects more low-energy modes in the MAM crystal given the spherical shape of monopoles compared to the drum-like shape of Skyrmions. It comes as a surprise that our LLG analysis revealed only one collective mode in low frequencies $\omega \lesssim J/\hbar$ (see SM).

Figure~\ref{fig:Fig2}(a) shows the real and imaginary parts of the Fourier spectrum $\chi_y (\omega) = \int_0^\infty e^{i\omega t} n_y (t)dt$, where $n_y (t)$ refers to the space-averaged $y$-component of the magnetization vector after the magnetic-field pulse $\delta B_y (t) = B_0 \delta (t)$ was applied along the $y$-direction, perpendicular to the external magnetic field. A single resonance peak at $\omega_{\rm R} \sim 0.03 J/\hbar$ was found. Other peaks in $\chi_y (\omega)$ exist, but at much higher frequencies $\omega > J/\hbar$. We found no low-energy absorption peaks in $\chi_z (\omega)$ when the pulse is applied along the Zeeman field, $\delta B_z (t) = B_0 \delta (t)$. With these observations we conclude that there exists only one low-frequency collective mode for the MAM crystal.

From the numerically obtained magnetization profile $\v n_{\v r} (t)$ we can keep track of the monopole positions and the Skyrmion density variations over time. First of all, the monopole and anti-monopole positions remained completely static over time either for a pulse perturbation $\delta \v B (t) = \v B_0 \delta (t)$ or for periodic perturbation $\delta \v B(t) = \v B_0 \sin (\omega_{\rm R} t)$ provided the strength $|\v B_0|$ is not too large to disturb the MAM crystal structure itself. Instead, layer-by-layer analysis of the Skyrmion density variation over time revealed some oscillatory behavior reminiscent of the Skyrmion's center-of-mass motion in the two-dimensional crystal. Real-time calculations for the two-dimensional Skyrmion lattice done in the past assumed a large Skyrmion radius \cite{mochizuki,Wang,Tokura_resonance}. The slow collective motion was easy to identify, typically by visual inspection of the real-time video generated from solving the LLG equation. Here we face a bigger challenge in the analysis due to the compactness of the topological object. With only about six lattice spacings between adjacent monopoles, the low-frequency motion becomes rather difficult to discern by inspection alone. The collective behavior must also be examined layer-by-layer as different $z$-layers (a plane in the cubic lattice sharing the same $z$-coordinate) might exhibit distinct dynamics. We have developed a characterization scheme based on the \emph{Skyrmion dipole vector}, defined as
\begin{align}
\v P (z, t) =\sum_{\v r_\perp = (x,y)}\Bigl(\rho^z_{\rm sk}(\v r_\perp, z, t)-\overline{ \rho^z_{\rm sk}(\v r_\perp, z, t) }\Bigr)\v r_\perp.\label{eq:sk_dipole}
\end{align}
Here $\rho^z_{\rm sk}(\v r_\perp, z, t)=(1/4\pi) \, \v n_{\v r} (t)  \cdot (\partial_x \v n_{\v r} (t) \times \partial_y \v n_{\v r} (t))$ refers to the Skyrmion density in the $xy$-plane at a given time $t$. As we know from elementary physics, the definition of the dipole moment suffers from ambiguity under the arbitrary translation of the lattice vector $\v r \rightarrow \v r + \v r_0$ ($\v r_0$ = arbitrary vector), unless the total ``charge'' is zero. To remedy this problem we subtract the time-averaged Skyrmion density $\overline{ \rho^z_{\rm sk}(\v r_\perp, z, t) }$, which is the average over several time periods of $\rho^z_{\rm sk}(\v r_\perp, z, t)$ for each lattice site $(\v r_\perp, z)$. Under a periodic magnetic-field perturbation $\delta B_y (t) = B_0 \sin (\omega_{\rm R} t)$, each $z$-layer develops a nonzero Skyrmion dipole vector $\v P (z, t)$, which oscillates with the resonant frequency $\omega_{\rm R}$. Figure~\ref{fig:Fig2}(b) shows the $\v P$-vector trajectories for different layers forming elliptic orbits of different orientations and sizes. The elongated shape of the orbit is due to the pulse applied in the $y$-direction that breaks the square lattice symmetry.  Note the centers of monopoles are located at the dual lattice sites, or between the layers of the original cubic lattice. In Fig.~\ref{fig:Fig2}(b) we plot the monopole and anti-monopole locations found from MC calculation for $L=30$ cubic lattice. It is seen that monopoles and anti-monopoles form between layers 1 and 2 of the original cubic lattice, and also between layers 4 and 5. The periodicity of the monopole lattice in the $z$-direction is 5, while it is 6 in the $xy$-plane (presumably due to the anisotropy induced by the Zeeman field). As shown in Fig.~\ref{fig:Fig2}(b), $\v P$-vectors for layers immediately above and immediately below the monopole layer oscillate at roughly right angles.

Both Skyrmions and monopoles are physical realizations of the integer-valued homotopy group $\pi_2 (S^2)$. Yet their properties differ in that the Skyrmion is a smooth texture in two spatial dimensions while the monopole is a singular object (at its center) embedded in three-dimensional space. Due to the singular nature of the monopole texture, it has been harder to find instances of its physical realization in condensed matter systems. Finally we are on the brink of claiming its realization and observation in a specific material MnGe \cite{tokura-MnGe-Hall,tokura-MnGe-Nernst,3D-imaging}, and understanding of their physical properties becomes of prominent importance. Existing theories are of Ginzburg-Landau nature \cite{Binz-Vishwanath,Park-Han} and cannot address dynamic properties. We have demonstrated here that a simple lattice model for chiral magnets, already known to yield the two-dimensional Skyrmion crystal phase at large helix periods, can also yield the monopole crystal phase by tuning the helix period to a smaller value. The low-energy collective behavior in the M-phase was examined numerically, with the conclusion that only one collective mode is observable at frequencies well below $J/\hbar$. Due to the finite-size constraint in MC simulations we can only tune the helix period discontinuously, and conclude that the period $\lambda=6$ definitely supports the monopole crystal phase even at low temperature. To compare with experiments, we adopt the known lattice constant 4.79~{\AA} for MnGe \cite{Mirebeau_NeutronMnGe} to translate the period $\lambda=6$ to 2.8~nm, which is slightly smaller than the experimentally determined spiral period of $3 \sim 6$~nm for this material. A search for candidate materials with even shorter helix period will be in favor of realizing the three-dimensional topological phase in a more robust manner.

Electronic structure coupled to the MAM crystal structure is expected to yield enormously interesting physics. One simple implication of the monopole formation is that it tends to trap a localized state of the electron \cite{syl}. Even more exciting possibility is whether the electronic system would develop some topological responses other than the Hall effect of Berry curvature origin \cite{witten-effect}.

We acknowledge Hye Jin Park and Jysoo Lee for helpful discussion. SGY and JHH are supported by NRF Grants No. 2013R1A2A1A01006430. YHL is supported by ERC Advanced Grant SIMCOFE. The supercomputing team at KISTI is gratefully acknowledged for providing computer resources and optimization support.

\bibliographystyle{apsrev}
\bibliography{reference.bib}

\begin{thebibliography}{27}
\expandafter\ifx\csname natexlab\endcsname\relax\def\natexlab#1{#1}\fi
\expandafter\ifx\csname bibnamefont\endcsname\relax
  \def\bibnamefont#1{#1}\fi
\expandafter\ifx\csname bibfnamefont\endcsname\relax
  \def\bibfnamefont#1{#1}\fi
\expandafter\ifx\csname citenamefont\endcsname\relax
  \def\citenamefont#1{#1}\fi
\expandafter\ifx\csname url\endcsname\relax
  \def\url#1{\texttt{#1}}\fi
\expandafter\ifx\csname urlprefix\endcsname\relax\def\urlprefix{URL }\fi
\providecommand{\bibinfo}[2]{#2}
\providecommand{\eprint}[2][]{\url{#2}}

\bibitem[{\citenamefont{Thomson}(1869)}]{thompson}
\bibinfo{author}{\bibfnamefont{W.}~\bibnamefont{Thomson}}, \bibinfo{journal}{P.
  Roy. Soc. Edin.} \textbf{\bibinfo{volume}{6}}, \bibinfo{pages}{94}
  (\bibinfo{year}{1869}).

\bibitem[{\citenamefont{Skyrme}(1962)}]{skyrme}
\bibinfo{author}{\bibfnamefont{T.}~\bibnamefont{Skyrme}},
  \bibinfo{journal}{Nucl. Phys.} \textbf{\bibinfo{volume}{31}},
  \bibinfo{pages}{556 } (\bibinfo{year}{1962}).

\bibitem[{\citenamefont{Mühlbauer et~al.}(2009)\citenamefont{Mühlbauer, Binz,
  Jonietz, Pfleiderer, Rosch, Neubauer, Georgii, and Böni}}]{pfleiderer09}
\bibinfo{author}{\bibfnamefont{S.}~\bibnamefont{Mühlbauer}},
  \bibinfo{author}{\bibfnamefont{B.}~\bibnamefont{Binz}},
  \bibinfo{author}{\bibfnamefont{F.}~\bibnamefont{Jonietz}},
  \bibinfo{author}{\bibfnamefont{C.}~\bibnamefont{Pfleiderer}},
  \bibinfo{author}{\bibfnamefont{A.}~\bibnamefont{Rosch}},
  \bibinfo{author}{\bibfnamefont{A.}~\bibnamefont{Neubauer}},
  \bibinfo{author}{\bibfnamefont{R.}~\bibnamefont{Georgii}}, \bibnamefont{and}
  \bibinfo{author}{\bibfnamefont{P.}~\bibnamefont{Böni}},
  \bibinfo{journal}{Science} \textbf{\bibinfo{volume}{323}},
  \bibinfo{pages}{915} (\bibinfo{year}{2009}).

\bibitem[{\citenamefont{Yi et~al.}(2009)\citenamefont{Yi, Onoda, Nagaosa, and
  Han}}]{YiSuDo}
\bibinfo{author}{\bibfnamefont{S.~D.} \bibnamefont{Yi}},
  \bibinfo{author}{\bibfnamefont{S.}~\bibnamefont{Onoda}},
  \bibinfo{author}{\bibfnamefont{N.}~\bibnamefont{Nagaosa}}, \bibnamefont{and}
  \bibinfo{author}{\bibfnamefont{J.~H.} \bibnamefont{Han}},
  \bibinfo{journal}{Phys. Rev. B} \textbf{\bibinfo{volume}{80}},
  \bibinfo{pages}{054416} (\bibinfo{year}{2009}).

\bibitem[{\citenamefont{Yu et~al.}(2010)\citenamefont{Yu, Onose, Kanazawa,
  Park, Han, Matsui, Nagaosa, and Tokura}}]{tokura10}
\bibinfo{author}{\bibfnamefont{X.~Z.} \bibnamefont{Yu}},
  \bibinfo{author}{\bibfnamefont{Y.}~\bibnamefont{Onose}},
  \bibinfo{author}{\bibfnamefont{N.}~\bibnamefont{Kanazawa}},
  \bibinfo{author}{\bibfnamefont{J.~H.} \bibnamefont{Park}},
  \bibinfo{author}{\bibfnamefont{J.~H.} \bibnamefont{Han}},
  \bibinfo{author}{\bibfnamefont{Y.}~\bibnamefont{Matsui}},
  \bibinfo{author}{\bibfnamefont{N.}~\bibnamefont{Nagaosa}}, \bibnamefont{and}
  \bibinfo{author}{\bibfnamefont{Y.}~\bibnamefont{Tokura}},
  \bibinfo{journal}{Nature} \textbf{\bibinfo{volume}{465}},
  \bibinfo{pages}{901} (\bibinfo{year}{2010}).

\bibitem[{\citenamefont{Heinze et~al.}(2011)\citenamefont{Heinze, von Bergmann,
  Menzel, Brede, Kubetzka, Wiesendanger, Bihlmayer, and Blugel}}]{wiesendanger}
\bibinfo{author}{\bibfnamefont{S.}~\bibnamefont{Heinze}},
  \bibinfo{author}{\bibfnamefont{K.}~\bibnamefont{von Bergmann}},
  \bibinfo{author}{\bibfnamefont{M.}~\bibnamefont{Menzel}},
  \bibinfo{author}{\bibfnamefont{J.}~\bibnamefont{Brede}},
  \bibinfo{author}{\bibfnamefont{A.}~\bibnamefont{Kubetzka}},
  \bibinfo{author}{\bibfnamefont{R.}~\bibnamefont{Wiesendanger}},
  \bibinfo{author}{\bibfnamefont{G.}~\bibnamefont{Bihlmayer}},
  \bibnamefont{and} \bibinfo{author}{\bibfnamefont{S.}~\bibnamefont{Blugel}},
  \bibinfo{journal}{Nat. Phys.} \textbf{\bibinfo{volume}{7}},
  \bibinfo{pages}{713} (\bibinfo{year}{2011}).

\bibitem[{\citenamefont{Seki et~al.}(2012)\citenamefont{Seki, Yu, Ishiwata, and
  Tokura}}]{tokura-MF-skyrmion}
\bibinfo{author}{\bibfnamefont{S.}~\bibnamefont{Seki}},
  \bibinfo{author}{\bibfnamefont{X.~Z.} \bibnamefont{Yu}},
  \bibinfo{author}{\bibfnamefont{S.}~\bibnamefont{Ishiwata}}, \bibnamefont{and}
  \bibinfo{author}{\bibfnamefont{Y.}~\bibnamefont{Tokura}},
  \bibinfo{journal}{Science} \textbf{\bibinfo{volume}{336}},
  \bibinfo{pages}{198} (\bibinfo{year}{2012}).

\bibitem[{\citenamefont{Nahas et~al.}(2015)\citenamefont{Nahas, Prokhorenko,
  Louis, Gui, Kornev, and Bellaiche}}]{ferroelectric_Sk}
\bibinfo{author}{\bibfnamefont{Y.}~\bibnamefont{Nahas}},
  \bibinfo{author}{\bibfnamefont{S.}~\bibnamefont{Prokhorenko}},
  \bibinfo{author}{\bibfnamefont{L.}~\bibnamefont{Louis}},
  \bibinfo{author}{\bibfnamefont{Z.}~\bibnamefont{Gui}},
  \bibinfo{author}{\bibfnamefont{I.}~\bibnamefont{Kornev}}, \bibnamefont{and}
  \bibinfo{author}{\bibfnamefont{L.}~\bibnamefont{Bellaiche}},
  \bibinfo{journal}{Nat. Commun.} \textbf{\bibinfo{volume}{6}}
  (\bibinfo{year}{2015}).

\bibitem[{\citenamefont{Han et~al.}(2010)\citenamefont{Han, Zang, Yang, Park,
  and Nagaosa}}]{han}
\bibinfo{author}{\bibfnamefont{J.~H.} \bibnamefont{Han}},
  \bibinfo{author}{\bibfnamefont{J.}~\bibnamefont{Zang}},
  \bibinfo{author}{\bibfnamefont{Z.}~\bibnamefont{Yang}},
  \bibinfo{author}{\bibfnamefont{J.-H.} \bibnamefont{Park}}, \bibnamefont{and}
  \bibinfo{author}{\bibfnamefont{N.}~\bibnamefont{Nagaosa}},
  \bibinfo{journal}{Phys. Rev. B} \textbf{\bibinfo{volume}{82}},
  \bibinfo{pages}{094429} (\bibinfo{year}{2010}).

\bibitem[{\citenamefont{Harada et~al.}(1992)\citenamefont{Harada, Matsuda,
  Bonevich, Igarashi, Kondo, Pozzi, Kawabe, and Tonomura}}]{tonomura}
\bibinfo{author}{\bibfnamefont{K.}~\bibnamefont{Harada}},
  \bibinfo{author}{\bibfnamefont{T.}~\bibnamefont{Matsuda}},
  \bibinfo{author}{\bibfnamefont{J.}~\bibnamefont{Bonevich}},
  \bibinfo{author}{\bibfnamefont{M.}~\bibnamefont{Igarashi}},
  \bibinfo{author}{\bibfnamefont{S.}~\bibnamefont{Kondo}},
  \bibinfo{author}{\bibfnamefont{G.}~\bibnamefont{Pozzi}},
  \bibinfo{author}{\bibfnamefont{U.}~\bibnamefont{Kawabe}}, \bibnamefont{and}
  \bibinfo{author}{\bibfnamefont{A.}~\bibnamefont{Tonomura}},
  \bibinfo{journal}{Nature} \textbf{\bibinfo{volume}{360}}, \bibinfo{pages}{51}
  (\bibinfo{year}{1992}).

\bibitem[{\citenamefont{Binz et~al.}(2006)\citenamefont{Binz, Vishwanath, and
  Aji}}]{Binz-Vishwanath}
\bibinfo{author}{\bibfnamefont{B.}~\bibnamefont{Binz}},
  \bibinfo{author}{\bibfnamefont{A.}~\bibnamefont{Vishwanath}},
  \bibnamefont{and} \bibinfo{author}{\bibfnamefont{V.}~\bibnamefont{Aji}},
  \bibinfo{journal}{Phys. Rev. Lett.} \textbf{\bibinfo{volume}{96}},
  \bibinfo{pages}{207202} (\bibinfo{year}{2006}).

\bibitem[{\citenamefont{Pfleiderer et~al.}(2004)\citenamefont{Pfleiderer,
  Reznik, Pintschovius, Lohneysen, Garst, and Rosch}}]{pfleiderer-Bragg-peak}
\bibinfo{author}{\bibfnamefont{C.}~\bibnamefont{Pfleiderer}},
  \bibinfo{author}{\bibfnamefont{D.}~\bibnamefont{Reznik}},
  \bibinfo{author}{\bibfnamefont{L.}~\bibnamefont{Pintschovius}},
  \bibinfo{author}{\bibfnamefont{H.~v.} \bibnamefont{Lohneysen}},
  \bibinfo{author}{\bibfnamefont{M.}~\bibnamefont{Garst}}, \bibnamefont{and}
  \bibinfo{author}{\bibfnamefont{A.}~\bibnamefont{Rosch}},
  \bibinfo{journal}{Nature} \textbf{\bibinfo{volume}{427}},
  \bibinfo{pages}{227} (\bibinfo{year}{2004}).

\bibitem[{\citenamefont{Park and Han}(2011)}]{Park-Han}
\bibinfo{author}{\bibfnamefont{J.-H.} \bibnamefont{Park}} \bibnamefont{and}
  \bibinfo{author}{\bibfnamefont{J.~H.} \bibnamefont{Han}},
  \bibinfo{journal}{Phys. Rev. B} \textbf{\bibinfo{volume}{83}},
  \bibinfo{pages}{184406} (\bibinfo{year}{2011}).

\bibitem[{\citenamefont{Kanazawa et~al.}(2011)\citenamefont{Kanazawa, Onose,
  Arima, Okuyama, Ohoyama, Wakimoto, Kakurai, Ishiwata, and
  Tokura}}]{tokura-MnGe-Hall}
\bibinfo{author}{\bibfnamefont{N.}~\bibnamefont{Kanazawa}},
  \bibinfo{author}{\bibfnamefont{Y.}~\bibnamefont{Onose}},
  \bibinfo{author}{\bibfnamefont{T.}~\bibnamefont{Arima}},
  \bibinfo{author}{\bibfnamefont{D.}~\bibnamefont{Okuyama}},
  \bibinfo{author}{\bibfnamefont{K.}~\bibnamefont{Ohoyama}},
  \bibinfo{author}{\bibfnamefont{S.}~\bibnamefont{Wakimoto}},
  \bibinfo{author}{\bibfnamefont{K.}~\bibnamefont{Kakurai}},
  \bibinfo{author}{\bibfnamefont{S.}~\bibnamefont{Ishiwata}}, \bibnamefont{and}
  \bibinfo{author}{\bibfnamefont{Y.}~\bibnamefont{Tokura}},
  \bibinfo{journal}{Phys. Rev. Lett.} \textbf{\bibinfo{volume}{106}},
  \bibinfo{pages}{156603} (\bibinfo{year}{2011}).

\bibitem[{\citenamefont{Shiomi et~al.}(2013)\citenamefont{Shiomi, Kanazawa,
  Shibata, Onose, and Tokura}}]{tokura-MnGe-Nernst}
\bibinfo{author}{\bibfnamefont{Y.}~\bibnamefont{Shiomi}},
  \bibinfo{author}{\bibfnamefont{N.}~\bibnamefont{Kanazawa}},
  \bibinfo{author}{\bibfnamefont{K.}~\bibnamefont{Shibata}},
  \bibinfo{author}{\bibfnamefont{Y.}~\bibnamefont{Onose}}, \bibnamefont{and}
  \bibinfo{author}{\bibfnamefont{Y.}~\bibnamefont{Tokura}},
  \bibinfo{journal}{Phys. Rev. B} \textbf{\bibinfo{volume}{88}},
  \bibinfo{pages}{064409} (\bibinfo{year}{2013}).

\bibitem[{\citenamefont{Tanigaki et~al.}(2015)\citenamefont{Tanigaki, Shibata,
  Kanazawa, Yu, Onose, Park, Shindo, and Tokura}}]{3D-imaging}
\bibinfo{author}{\bibfnamefont{T.}~\bibnamefont{Tanigaki}},
  \bibinfo{author}{\bibfnamefont{K.}~\bibnamefont{Shibata}},
  \bibinfo{author}{\bibfnamefont{N.}~\bibnamefont{Kanazawa}},
  \bibinfo{author}{\bibfnamefont{X.}~\bibnamefont{Yu}},
  \bibinfo{author}{\bibfnamefont{Y.}~\bibnamefont{Onose}},
  \bibinfo{author}{\bibfnamefont{H.~S.} \bibnamefont{Park}},
  \bibinfo{author}{\bibfnamefont{D.}~\bibnamefont{Shindo}}, \bibnamefont{and}
  \bibinfo{author}{\bibfnamefont{Y.}~\bibnamefont{Tokura}},
  \bibinfo{journal}{Nano Lett.} \textbf{\bibinfo{volume}{15}},
  \bibinfo{pages}{5438} (\bibinfo{year}{2015}).

\bibitem[{\citenamefont{Buhrandt and Fritz}(2013)}]{Lars}
\bibinfo{author}{\bibfnamefont{S.}~\bibnamefont{Buhrandt}} \bibnamefont{and}
  \bibinfo{author}{\bibfnamefont{L.}~\bibnamefont{Fritz}},
  \bibinfo{journal}{Phys. Rev. B} \textbf{\bibinfo{volume}{88}},
  \bibinfo{pages}{195137} (\bibinfo{year}{2013}).

\bibitem[{\citenamefont{Yu et~al.}(2011)\citenamefont{Yu, Kanazawa, Onose,
  Kimoto, Zhang, Ishiwata, Matsui, and Tokura}}]{tokura_thinfilm}
\bibinfo{author}{\bibfnamefont{X.~Z.} \bibnamefont{Yu}},
  \bibinfo{author}{\bibfnamefont{N.}~\bibnamefont{Kanazawa}},
  \bibinfo{author}{\bibfnamefont{Y.}~\bibnamefont{Onose}},
  \bibinfo{author}{\bibfnamefont{K.}~\bibnamefont{Kimoto}},
  \bibinfo{author}{\bibfnamefont{W.~Z.} \bibnamefont{Zhang}},
  \bibinfo{author}{\bibfnamefont{S.}~\bibnamefont{Ishiwata}},
  \bibinfo{author}{\bibfnamefont{Y.}~\bibnamefont{Matsui}}, \bibnamefont{and}
  \bibinfo{author}{\bibfnamefont{Y.}~\bibnamefont{Tokura}},
  \bibinfo{journal}{Nat. Mater.} \textbf{\bibinfo{volume}{10}},
  \bibinfo{pages}{106} (\bibinfo{year}{2011}).

\bibitem[{\citenamefont{Mochizuki}(2012)}]{mochizuki}
\bibinfo{author}{\bibfnamefont{M.}~\bibnamefont{Mochizuki}},
  \bibinfo{journal}{Phys. Rev. Lett.} \textbf{\bibinfo{volume}{108}},
  \bibinfo{pages}{017601} (\bibinfo{year}{2012}).

\bibitem[{\citenamefont{Liu et~al.}(2013)\citenamefont{Liu, Li, and
  Han}}]{yehua_han}
\bibinfo{author}{\bibfnamefont{Y.-H.} \bibnamefont{Liu}},
  \bibinfo{author}{\bibfnamefont{Y.-Q.} \bibnamefont{Li}}, \bibnamefont{and}
  \bibinfo{author}{\bibfnamefont{J.~H.} \bibnamefont{Han}},
  \bibinfo{journal}{Phys. Rev. B} \textbf{\bibinfo{volume}{87}},
  \bibinfo{pages}{100402} (\bibinfo{year}{2013}).

\bibitem[{\citenamefont{Lin et~al.}(2014)\citenamefont{Lin, Batista, and
  Saxena}}]{Batista}
\bibinfo{author}{\bibfnamefont{S.-Z.} \bibnamefont{Lin}},
  \bibinfo{author}{\bibfnamefont{C.~D.} \bibnamefont{Batista}},
  \bibnamefont{and} \bibinfo{author}{\bibfnamefont{A.}~\bibnamefont{Saxena}},
  \bibinfo{journal}{Phys. Rev. B} \textbf{\bibinfo{volume}{89}},
  \bibinfo{pages}{024415} (\bibinfo{year}{2014}).

\bibitem[{\citenamefont{Wang et~al.}(2015)\citenamefont{Wang, Beg, Zhang, Kuch,
  and Fangohr}}]{Wang}
\bibinfo{author}{\bibfnamefont{W.}~\bibnamefont{Wang}},
  \bibinfo{author}{\bibfnamefont{M.}~\bibnamefont{Beg}},
  \bibinfo{author}{\bibfnamefont{B.}~\bibnamefont{Zhang}},
  \bibinfo{author}{\bibfnamefont{W.}~\bibnamefont{Kuch}}, \bibnamefont{and}
  \bibinfo{author}{\bibfnamefont{H.}~\bibnamefont{Fangohr}},
  \bibinfo{journal}{Phys. Rev. B} \textbf{\bibinfo{volume}{92}},
  \bibinfo{pages}{020403} (\bibinfo{year}{2015}).

\bibitem[{\citenamefont{Okamura et~al.}(2013)\citenamefont{Okamura, Kagawa,
  Mochizuki, Kubota, Seki, Ishiwata, Kawasaki, Onose, and
  Tokura}}]{Tokura_resonance}
\bibinfo{author}{\bibfnamefont{Y.}~\bibnamefont{Okamura}},
  \bibinfo{author}{\bibfnamefont{F.}~\bibnamefont{Kagawa}},
  \bibinfo{author}{\bibfnamefont{M.}~\bibnamefont{Mochizuki}},
  \bibinfo{author}{\bibfnamefont{M.}~\bibnamefont{Kubota}},
  \bibinfo{author}{\bibfnamefont{S.}~\bibnamefont{Seki}},
  \bibinfo{author}{\bibfnamefont{S.}~\bibnamefont{Ishiwata}},
  \bibinfo{author}{\bibfnamefont{M.}~\bibnamefont{Kawasaki}},
  \bibinfo{author}{\bibfnamefont{Y.}~\bibnamefont{Onose}}, \bibnamefont{and}
  \bibinfo{author}{\bibfnamefont{Y.}~\bibnamefont{Tokura}},
  \bibinfo{journal}{Nat. Commun.} \textbf{\bibinfo{volume}{4}}
  (\bibinfo{year}{2013}).

\bibitem[{\citenamefont{Onose et~al.}(2012)\citenamefont{Onose, Okamura, Seki,
  Ishiwata, and Tokura}}]{Tokura_observation}
\bibinfo{author}{\bibfnamefont{Y.}~\bibnamefont{Onose}},
  \bibinfo{author}{\bibfnamefont{Y.}~\bibnamefont{Okamura}},
  \bibinfo{author}{\bibfnamefont{S.}~\bibnamefont{Seki}},
  \bibinfo{author}{\bibfnamefont{S.}~\bibnamefont{Ishiwata}}, \bibnamefont{and}
  \bibinfo{author}{\bibfnamefont{Y.}~\bibnamefont{Tokura}},
  \bibinfo{journal}{Phys. Rev. Lett.} \textbf{\bibinfo{volume}{109}},
  \bibinfo{pages}{037603} (\bibinfo{year}{2012}).

\bibitem[{\citenamefont{Makarova et~al.}(2012)\citenamefont{Makarova,
  Tsvyashchenko, Andre, Porcher, Fomicheva, Rey, and
  Mirebeau}}]{Mirebeau_NeutronMnGe}
\bibinfo{author}{\bibfnamefont{O.~L.} \bibnamefont{Makarova}},
  \bibinfo{author}{\bibfnamefont{A.~V.} \bibnamefont{Tsvyashchenko}},
  \bibinfo{author}{\bibfnamefont{G.}~\bibnamefont{Andre}},
  \bibinfo{author}{\bibfnamefont{F.}~\bibnamefont{Porcher}},
  \bibinfo{author}{\bibfnamefont{L.~N.} \bibnamefont{Fomicheva}},
  \bibinfo{author}{\bibfnamefont{N.}~\bibnamefont{Rey}}, \bibnamefont{and}
  \bibinfo{author}{\bibfnamefont{I.}~\bibnamefont{Mirebeau}},
  \bibinfo{journal}{Phys. Rev. B} \textbf{\bibinfo{volume}{85}},
  \bibinfo{pages}{205205} (\bibinfo{year}{2012}).

\bibitem[{\citenamefont{Lee and Han}(2015)}]{syl}
\bibinfo{author}{\bibfnamefont{S.-Y.} \bibnamefont{Lee}} \bibnamefont{and}
  \bibinfo{author}{\bibfnamefont{J.~H.} \bibnamefont{Han}},
  \bibinfo{journal}{Phys. Rev. B} \textbf{\bibinfo{volume}{91}},
  \bibinfo{pages}{245121} (\bibinfo{year}{2015}).

\bibitem[{\citenamefont{Watanabe and Vishwanath}(2014)}]{witten-effect}
\bibinfo{author}{\bibfnamefont{H.}~\bibnamefont{Watanabe}} \bibnamefont{and}
  \bibinfo{author}{\bibfnamefont{A.}~\bibnamefont{Vishwanath}},
  \bibinfo{journal}{arXiv:1410.2213}  (\bibinfo{year}{2014}).

\end{thebibliography}


\begin{thebibliography}{5}
\expandafter\ifx\csname natexlab\endcsname\relax\def\natexlab#1{#1}\fi
\expandafter\ifx\csname bibnamefont\endcsname\relax
  \def\bibnamefont#1{#1}\fi
\expandafter\ifx\csname bibfnamefont\endcsname\relax
  \def\bibfnamefont#1{#1}\fi
\expandafter\ifx\csname citenamefont\endcsname\relax
  \def\citenamefont#1{#1}\fi
\expandafter\ifx\csname url\endcsname\relax
  \def\url#1{\texttt{#1}}\fi
\expandafter\ifx\csname urlprefix\endcsname\relax\def\urlprefix{URL }\fi
\providecommand{\bibinfo}[2]{#2}
\providecommand{\eprint}[2][]{\url{#2}}

\bibitem[{\citenamefont{Buhrandt and Fritz}(2013)}]{Lars}
\bibinfo{author}{\bibfnamefont{S.}~\bibnamefont{Buhrandt}} \bibnamefont{and}
  \bibinfo{author}{\bibfnamefont{L.}~\bibnamefont{Fritz}},
  \bibinfo{journal}{Phys. Rev. B} \textbf{\bibinfo{volume}{88}},
  \bibinfo{pages}{195137} (\bibinfo{year}{2013}).

\bibitem[{\citenamefont{Preis et~al.}(2009)\citenamefont{Preis, Virnau, Paul,
  and Schneider}}]{checkerboard}
\bibinfo{author}{\bibfnamefont{T.}~\bibnamefont{Preis}},
  \bibinfo{author}{\bibfnamefont{P.}~\bibnamefont{Virnau}},
  \bibinfo{author}{\bibfnamefont{W.}~\bibnamefont{Paul}}, \bibnamefont{and}
  \bibinfo{author}{\bibfnamefont{J.~J.} \bibnamefont{Schneider}},
  \bibinfo{journal}{J. Comput. Phys.} \textbf{\bibinfo{volume}{228}},
  \bibinfo{pages}{4468 } (\bibinfo{year}{2009}).

\bibitem[{\citenamefont{Milde et~al.}(2013)\citenamefont{Milde, Köhler,
  Seidel, Eng, Bauer, Chacon, Kindervater, Mühlbauer, Pfleiderer, Buhrandt
  et~al.}}]{pfleiderer-monopole}
\bibinfo{author}{\bibfnamefont{P.}~\bibnamefont{Milde}},
  \bibinfo{author}{\bibfnamefont{D.}~\bibnamefont{Köhler}},
  \bibinfo{author}{\bibfnamefont{J.}~\bibnamefont{Seidel}},
  \bibinfo{author}{\bibfnamefont{L.~M.} \bibnamefont{Eng}},
  \bibinfo{author}{\bibfnamefont{A.}~\bibnamefont{Bauer}},
  \bibinfo{author}{\bibfnamefont{A.}~\bibnamefont{Chacon}},
  \bibinfo{author}{\bibfnamefont{J.}~\bibnamefont{Kindervater}},
  \bibinfo{author}{\bibfnamefont{S.}~\bibnamefont{Mühlbauer}},
  \bibinfo{author}{\bibfnamefont{C.}~\bibnamefont{Pfleiderer}},
  \bibinfo{author}{\bibfnamefont{S.}~\bibnamefont{Buhrandt}},
  \bibnamefont{et~al.}, \bibinfo{journal}{Science}
  \textbf{\bibinfo{volume}{340}}, \bibinfo{pages}{1076} (\bibinfo{year}{2013}).

\bibitem[{\citenamefont{Sch\"utte and Rosch}(2014)}]{rosch-monopole}
\bibinfo{author}{\bibfnamefont{C.}~\bibnamefont{Sch\"utte}} \bibnamefont{and}
  \bibinfo{author}{\bibfnamefont{A.}~\bibnamefont{Rosch}},
  \bibinfo{journal}{Phys. Rev. B} \textbf{\bibinfo{volume}{90}},
  \bibinfo{pages}{174432} (\bibinfo{year}{2014}).

\bibitem[{\citenamefont{van Oosterom and Strackee}(1983)}]{solid_angle}
\bibinfo{author}{\bibfnamefont{A.}~\bibnamefont{van Oosterom}}
  \bibnamefont{and} \bibinfo{author}{\bibfnamefont{J.}~\bibnamefont{Strackee}},
  \bibinfo{journal}{Biomedical Engineering, IEEE Transactions on}
  \textbf{\bibinfo{volume}{BME-30}}, \bibinfo{pages}{125}
  (\bibinfo{year}{1983}).

\end{thebibliography}

\end{document}


\title{Supplemental Material for ``Formation of a Topological Monopole Lattice and its Dynamics in Three-dimensional Chiral Magnets''}
\author{Seong-Gyu Yang}
\affiliation{Department of Physics, Sungkyunkwan University, Suwon 16419, Korea}
\author{Ye-Hua Liu}
\affiliation{Theoretische Physik, ETH Zurich, 8093 Zurich, Switzerland}
\author{Jung Hoon Han}\email{hanjh@skku.edu}
\affiliation{Department of Physics, Sungkyunkwan University, Suwon 16419, Korea}

\date{\today}

\maketitle

\section{GPU-based Monte Carlo simulations}
We adopt the Buhrandt-Fritz model \cite{Lars} including the ferromagnetic and DM interactions for nearest and next-nearest neighbors, and the Zeeman term:
\begin{eqnarray}
\mathcal{H} =
&& -J\sum_{\v r}\v{n}_{\v r}\cdot\bigr(\v{n}_{\v r+\hat x}+\v{n}_{\v r+\hat y}+\v{n}_{\v r+\hat z} \bigl) \nn
&& +\frac{J}{16}\sum_{\v r}\v{n}_{\v r}\cdot\bigr(\v{n}_{\v r+2\hat x}+\v{n}_{\v r+2\hat y}+\v{n}_{\v r+2\hat z} \bigl) \nn
&& -D \sum_{\v r} \v{n}_{\v r} \cdot\bigr(\v{n}_{\v r+\hat x}\times\hat x + \v{n}_{\v r+\hat y} \times \hat y + \v{n}_{\v r+\hat z} \times \hat z \bigl) \nn
&& +\frac{D}{8} \sum_{\v r}\v{n}_{\v r} \cdot\bigr(\v{n}_{\v r+2\hat x}\times\hat x + \v{n}_{\v r+2\hat y} \times \hat y + \v{n}_{\v r+2\hat z} \times \hat z \bigl) \nn
&& -\v B \cdot \sum_{\v r}\v n_{\v r}. \label{eq:model-H}
\end{eqnarray}
Each spin $\v n_{\v r}$ is a classical unit vector defined at the sites of the cubic lattice. The external magnetic field $\v B$ is applied along the $z$-direction. The ground state of this model at $\v B=0$ has the (111) spiral structure with the wave number $k=|\v k|/\sqrt{3}$ for $x$, $y$ and $z$ direction given by the equation
\begin{align}
\frac{D}{J} = \frac{\sqrt{3}\sin k (4-\cos k)}{4\cos k -\cos2k}. \label{eq:D-over-J}
\end{align}
We use this equation to determine the ratio $D/J$ for the desired spiral period $\lambda = 2\pi/k$.

All Monte Carlo calculations were performed on a cubic lattice of size $L^3$ with periodic boundary conditions. $L$ was chosen to be commensurate with the expected helix period. In particular we choose $L=30$ for $\lambda=6,10$ and $L=32$ for $\lambda=8$. Following the previous work \cite{Lars}, three different kinds of MC simulations were performed in order to obtain the thermodynamically stable state. They are the fixed-field annealing, and the fixed-temperature field-decreasing and field-increasing simulations. In the fixed-field annealing simulation, the magnetic field range was divided into 30 steps from $B=3J$ to $B=0$. At each field, the temperature was successively lowered over 40 annealing steps for $\lambda=8, 10$, and 160 steps for $\lambda=6$, ranging from $T=4.01J$ to $T=0.01J$. Random initial configurations were adopted at the highest temperature. For the fixed-temperature, field-decreasing (increasing) simulation, we started with spin configurations obtained from the earlier fixed-field annealing procedure at high (low) field and gradually lowered (raised) the external magnetic field over 30 steps. After all the simulations were finished, we identified the state with the lowest free energy among the three annealing processes, for every temperature and field values.

MC simulations for three-dimensional lattices are very computer-intensive. Instead of running the program on individual central processing units (CPUs) handling less than 10 threads per processor, we used the general-purpose computing on graphics processing unit (GPGPU) method, utilizing the Compute Unified Device Architecture (CUDA). In this method, several thousands of threads could run at the same time. Synchronization problem occurs in the parallelization of the MC calculation among the threads. When one thread is proposing a local MC update for a spin, other threads, running at the same time, might change the surrounding spins. The exchange energy obtained in this way is not guaranteed to be correct. This problem can be avoided using the two-color checkerboard algorithm \cite{checkerboard}, if there were only nearest-neighbor interactions. To handle next-nearest-neighbor interactions in Eq.~(\ref{eq:model-H}), we used a three-color algorithm, wherein each lattice site with coordinates $(i_x,i_y,i_z)$ ($0\le i_x,i_y,i_z \le L-1$), is colored by the remainder $c=(i_x+i_y+i_z) \bmod 3$. During one MC step, we performed parallel MC simulations in the order of $c=0$, $c=1$, and $c=2$. All threads were synchronized after simulations of all three colors.

\section{Monopole number calculation}
The topological charge density of Skyrmions in two-dimensional magnets is measured by $\rho_{\rm sk} = (1/4\pi) \, \v n \cdot (\partial_x \v n \times \partial_y \v n)$. This is generalized in three-dimensional magnets as a vector of Skyrmion densities,
\begin{align}
\rho^\alpha_{\rm sk} = \frac{1}{8\pi} \varepsilon^{\alpha\beta\gamma} \v n \!\cdot \!(\partial_\beta \v n \!\times \!\partial_\gamma \v n) ,\label{eq:monople_density}
\end{align}
where each component measures the Skyrmion density within the $(\beta\gamma)$-plane normal to the given direction $\alpha=x,y,z$. The divergence of the vector $\rho_{\rm m} = \bm \nabla \cdot \bm \rho_{\rm sk}$ then measures the monopole density in the magnetic structure \cite{pfleiderer-monopole,rosch-monopole}. For the typical monopole and anti-monopole configurations $\v n_{\v r} = \pm \v r /r$, one obtains $ \rho_{\rm m} = \pm \delta (\v r)$ by explicit calculation.

A suitable discretization of the divergence $\bm \nabla \cdot \bm \rho_{\rm sk}$ on a lattice is necessary to read off the monopole locations from a given magnetization structure. In the cubic lattice, each elementary cube  consists of 8 spins. Each face of the cube is divided into two triangles with three spins for each triangle. Solid angles subtended by each set of three spins were calculated using Oosterom's method \cite{solid_angle}. Finally, all solid angles from the six faces of the cube, twelve triangles in total, were summed and divided by $4\pi$ to give the monopole number contained in this cube. The resulting monopole number is either 0, +1 (monopole), or -1 (anti-monopole). No other number was ever found from the spin structure. Monopole locations are easily identified as the cubes with nonzero integers.

\section{Low-frequency monopole lattice dynamics}
To study the time-dependent spin dynamics excited by a magnetic-field pulse, we numerically solve the Landau-Lifshitz-Gilbert (LLG) equation
\begin{align}
\dot{\v n}_\v r=\frac{1}{1+\alpha^2}\Bigl[ \frac{\delta\cal H}{\delta\v{n}_\v r}  \times \v{n}_\v r + \alpha\v{n}_\v r \times \Bigl(\v{n}_\v r \times \frac{\delta\cal H}{\delta\v{n}_\v r}\Bigr) \Bigr],
\label{eq:LLG}
\end{align}
for the $24^3$ and $30^3$ cubic lattices in the M-phase, using the fourth-order Runge-Kutta method. Here, $\alpha=0.04$ is the Gilbert damping constant and we choose the time unit $t_0 = \hbar/J$ to make Eq.~(\ref{eq:LLG}) dimensionless. An additional magnetic-field pulse $\v B_0\delta(t)$ was applied initially in either $y$ or $z$ direction to excite the MAM lattice. The subsequent time evolution is obtained by solving the LLG equation. The Fast Fourier Transform (FFT) of the magnetization curve along the $y$-direction is used to calculate the ac susceptibility $\chi_y$, shown as Fig.~2(a) in the main text.

The pulse applied in the $z$-direction, coinciding with the external magnetic field orientation, produces no low-frequency responses in the crystal. The $y$-direction pulse, on the other hand, produces a single peak at $\omega_\mathrm{R} \sim 0.03J/\hbar$. We then applied a sinusoidal magnetic field $B_y (t) = B_0 \sin ( \omega_\mathrm{R} t )$ and examined the time evolution of the magnetization vector $\v n_{\v r} (t)$. Once $\v n_{\v r} (t)$ is obtained, we compute the Skyrmion dipole vector according to Eq.~(2) in the main text.

\bibliographystyle{apsrev}
\bibliography{reference.bib}